\documentclass[%
 reprint,
superscriptaddress,
 amsmath,amssymb,
 aps,
]{revtex4-1}

\usepackage{graphicx}
\usepackage{color}
\usepackage{dcolumn}
\usepackage{bm}
\usepackage{ amssymb }

\newcommand{\bologna}{\affiliation{Department of Physics and Astrophysics, University of Bologna and INFN-Bologna, 40126 Bologna, Italy}}
\newcommand{\chicago}{\affiliation{Department of Physics \& Kavli Institute of Cosmological Physics, University of Chicago, Chicago, IL 60637, USA}}
\newcommand{\coimbra}{\affiliation{LIBPhys, Department of Physics, University of Coimbra, 3004-516 Coimbra, Portugal}}
\newcommand{\columbia}{\affiliation{Physics Department, Columbia University, New York, NY 10027, USA}}
\newcommand{\lngs}{\affiliation{INFN-Laboratori Nazionali del Gran Sasso and Gran Sasso Science Institute, 67100 L'Aquila, Italy}}
\newcommand{\freiburg}{\affiliation{Physikalisches Institut, Universit\"at Freiburg, 79104 Freiburg, Germany}}
\newcommand{\mainz}{\affiliation{Institut f\"ur Physik \& Exzellenzcluster PRISMA, Johannes Gutenberg-Universit\"at Mainz, 55099 Mainz, Germany}}
\newcommand{\heidelberg}{\affiliation{Max-Planck-Institut f\"ur Kernphysik, 69117 Heidelberg, Germany}}
\newcommand{\munster}{\affiliation{Institut f\"ur Kernphysik, Westf\"alische Wilhelms-Universit\"at M\"unster, 48149 M\"unster, Germany}}
\newcommand{\nikhef}{\affiliation{Nikhef and the University of Amsterdam, Science Park, 1098XG Amsterdam, Netherlands}}
\newcommand{\nyuad}{\affiliation{New York University Abu Dhabi, Abu Dhabi, United Arab Emirates}}
\newcommand{\paris}{\affiliation{LPNHE, Universit\'e Pierre et Marie Curie, Universit\'e Paris Diderot, CNRS/IN2P3, Paris 75252, France}}
\newcommand{\purdue}{\affiliation{Department of Physics and Astronomy, Purdue University, West Lafayette, IN 47907, USA}}
\newcommand{\rpi}{\affiliation{Department of Physics, Applied Physics and Astronomy, Rensselaer Polytechnic Institute, Troy, NY 12180, USA}}
\newcommand{\rice}{\affiliation{Department of Physics and Astronomy, Rice University, Houston, TX 77005, USA}}
\newcommand{\stockholm}{\affiliation{Oskar Klein Centre, Department of Physics, Stockholm University, AlbaNova, Stockholm SE-10691, Sweden}}
\newcommand{\subatech}{\affiliation{SUBATECH, IMT Atlantique, CNRS/IN2P3, Universit\'e de Nantes, Nantes 44307, France}}
\newcommand{\torino}{\affiliation{INFN-Torino and Osservatorio Astrofisico di Torino, 10125 Torino, Italy}}
\newcommand{\ucla}{\affiliation{Physics \& Astronomy Department, University of California, Los Angeles, CA 90095, USA}}
\newcommand{\ucsd}{\affiliation{Department of Physics, University of California, San Diego, CA 92093, USA}}
\newcommand{\wis}{\affiliation{Department of Particle Physics and Astrophysics, Weizmann Institute of Science, Rehovot 7610001, Israel}}
\newcommand{\zurich}{\affiliation{Physik-Institut, University of Zurich, 8057  Zurich, Switzerland}}

\begin{document}


\title{Search for Bosonic Super-WIMP Interactions with the XENON100 Experiment}

\author{E.~Aprile}\columbia
\author{J.~Aalbers}\nikhef
\author{F.~Agostini}\lngs\bologna
\author{M.~Alfonsi}\mainz
\author{L.~Althueser}\munster
\author{F.~D.~Amaro}\coimbra
\author{M.~Anthony}\columbia
\author{F.~Arneodo}\nyuad
\author{P.~Barrow}\zurich
\author{L.~Baudis}\email{laura.baudis@uzh.ch}\zurich
% ...
\collaboration{XENON Collaboration}\email{xenon@lngs.infn.it}\noaffiliation

\date{\today}

\begin{abstract}
We present results of searches for vector and pseudo-scalar bosonic super-WIMPs, which are dark matter candidates with masses at the keV-scale, with the XENON100 experiment. XENON100 is a dual-phase xenon time projection chamber operated at the Laboratori Nazionali del Gran Sasso. A profile likelihood analysis of data with an exposure of 224.6 live days $\times$ 34\,kg showed no evidence for a signal above the expected background. We thus obtain new and stringent upper limits in the $(8-125)$\,keV/c$^2$ mass range, excluding couplings to electrons with coupling constants of $g_{ae} > 3\times10^{-13}$  for pseudo-scalar and $\alpha'/\alpha > 2\times10^{-28}$ for vector  super-WIMPs,  respectively. These limits are derived under the assumption that super-WIMPs constitute all of the dark matter in our galaxy.

\end{abstract}

\maketitle

\section{\label{sec:intro} Introduction}

There is overwhelming evidence for the presence of dark matter in our universe. Its existence is inferred from a variety of observations, including those of the temperature fluctuations in the cosmic microwave background \cite{Ade:2013zuv}, gravitational lensing \cite{Kaiser:1992ps}, mass-to-light ratio in galaxy clusters \cite{Zwicky:1933gu}, and galactic rotation curves \cite{Rubin:1970zza}. In addition, simulations of large-scale structure and galaxy formation require the presence of non-baryonic matter to reproduce the observed cosmic structures~\cite{Frenk:2012ph}.

While the microscopic nature of dark matter is largely unknown, the simplest assumption which can explain all existing observations is that it is made of a new, yet undiscovered particle~\cite{Bertone:2010zza}.  Leading examples are Weakly Interacting Massive Particles (WIMPs), axions or axion-like-particles (ALPs) and sterile neutrinos. WIMPs with masses in the GeV range, as well as axions/ALPs are examples for Cold Dark Matter (CDM), while sterile neutrinos with masses at the keV-scale are an example for Warm Dark Matter (WDM). While CDM particles were non-relativistic at the time of their decoupling from the rest of the particles in the early universe, WDM particles remain relativistic for longer, retaining a larger velocity dispersion and more easily free-streaming out from small-scale perturbations. Astrophysical and cosmological observations constrain the mass of WDM to be larger than $\sim$3\,keV/c$^2$~\cite{Markovic:2013iza,Lapi:2015zea}, with a more recent lower limit from Lyman-$\alpha$ forest data being 5.3\,keV \cite{Irsic:2017ixq}.

A large number of experiments aim to observe axions/ALPs and WIMPs: directly, indirectly, or via production at the LHC \cite{Baudis:2015mpa}. Direct detection experiments, which look for low energy nuclear recoils produced in collisions of WIMPs with atomic nuclei feature low energy thresholds, large detector masses and ultra-low backgrounds~\cite{Undagoitia:2015gya, Baudis:2016qwx}. Such experiments can thus also observe other type of particles, with non-vanishing couplings to electrons. Among these, bosonic super-WIMPs~\cite{Pospelov:2008jk} are an example for WDM. These particles, with masses at the keV-scale, could couple electromagnetically to Standard Model particles via the axio-electric effect, which  is an analogous process  to the photoelectric effect, and thus be detected in direct detection experiments~\cite{Pospelov:2008jk}.

In this study we present a search for vector and pseudoscalar bosonic super-WIMPs with the XENON100 detector.  The super-WIMPs can be absorbed in liquid xenon  (LXe) and the expected signature is a mono-energetic peak at the super-WIMP's rest mass.  We have presented first results on pseudoscalar bosonic super-WIMPs, or axion-like particles, in ref.~\cite{Aprile:2014eoa}. In this analysis we include two major improvements: we extend the mass range up to 125\,keV/c$^2$ and we improve the energy resolution by employing both scintillation and ionisation signals to determine the energy scale and resolution of the XENON100 detector.

This paper is organised as follows. We describe the main features of the XENON100 experiment in section~\ref{sec:xenon100}, after which we briefly review the expected signal and rates in a dark matter detector in section~\ref{sec:signal}.  We detail the data analysis methods in section~\ref{sec:analysis} and present our main findings in section~\ref{sec:results}, together with a  discussion of their implications.

\section{\label{sec:xenon100} The XENON100 Detector}

Located at the Laboratori Nazionali del Gran Sasso (LNGS), the XENON100 experiment operates a dual-phase (liquid and gas) xenon time projection chamber (TPC). The detector contains 161\,kg of LXe in total,  with  62\,kg in the active region of the TPC. A total of 178 1-inch square, low-radioactivity, UV-sensitive photo-multiplier tubes (PMTs) arranged in two arrays, one in the liquid and one in the gas, detect the prompt scintillation (S1) and the delayed, proportional scintillation signal (S2) 
arising when a particle interacts in the TPC. The ionisation electrons are  drifted via an electric field of 530\,V/cm to the liquid-gas boundary and extracted in the vapour phase via a $\sim$12\,kV/cm field, where  proportional scintillation is produced.
 The three-dimensional position  of the original interaction site is reconstructed via the time difference between the S1 and S2 signal ($z$-position, with a 1\,$\sigma$ resolution $<$0.3\,mm), and by exploiting the S2 light pattern in the top PMT array ($(x,y)$-position, with a 1\,$\sigma$ resolution $<$3\,mm). This enables us to reject a large fraction of background events via fiducial volume cuts and selection of single-scatters  \cite{Aprile:2011dd}. A 4\,cm thick, 99\,kg LXe, layer surrounds the TPC and is observed by 64 1-inch square PMTs, allowing us to rejects events with energy depositions in this active LXe region. Finally,  nuclear recoils (NRs) induced by fast neutrons and WIMP-nucleus scatters, and electronic recoils (ERs), produced by $\beta$ and $\gamma$-rays, as well as axions, can be distinguished based on their (S2/S1)-ratio.

The XENON100 experiment, described in detail in~\cite{Aprile:2011dd}, has delivered competitive constraints on spin-independent~\cite{Aprile:2012nq,Aprile:2016swn} and spin-dependent~\cite{Aprile:2013doa,Aprile:2016swn} WIMP-nucleus 
scatters, on solar axions and galactic ALPs~\cite{Aprile:2014eoa}, as well as on leptophilic dark matter models~\cite{Aprile:2015ade,Aprile:2015ibr,Aprile:2017yea}.  In this analysis, we use the Run\,II science data, with 224.6~live days of data taking and 34\,kg of LXe in the fiducial region, to search for signatures from bosonic super-WIMPs in the mass range $(8-125)$\,keV/c$^2$ in the ER spectrum. A previous search in LXe was conducted by XMASS, constraining the mass range $(40-120)$\,keV/c$^2$~\cite{Abe:2014zcd}.

\section{\label{sec:signal}Expected signal}

Viable models for super-WIMPs as dark matter candidates result in vector and pseudo-scalar particles~\cite{Pospelov:2008jk}. The expected interaction rates in a detector are obtained by convoluting the absorption cross section with the expected flux of these particles. The absorption cross section $\sigma_{abs}$  for vector super-WIMPs can be written in terms of the cross section for photon absorption via the photoelectric effect $\sigma_{pe}$, with the photon energy $\omega$ replaced by the mass of the vector boson m$_{v}$:

\begin{equation}
\label{eq:vecscale}
\frac{\sigma_{abs}v}{\sigma_{pe}(\omega=m_{v})c}\simeq\frac{\alpha'}{\alpha},  
\end{equation}

\noindent
where  $v$ is the incoming velocity of the  vector boson, $c$ is the velocity of light, $\alpha$ and $\alpha'$ are the fine structure constant, and its vector boson equivalent, respectively.

For pseudo-scalar super-WIMPs, the relation between the two cross sections is as follows:
\begin{equation}
\label{eq:scascale}
\frac{\sigma_{abs}v}{\sigma_{pe}(\omega=m_{a})c}\simeq\frac{3m_{a}^{2}}{4 \pi \alpha f_{a}^{2}}, 
\end{equation}

\noindent
where $m_{a}$ is the mass of the pseudo-scalar particle, and $f_{a}$ is a dimensional coupling constant.

Assuming that super-WIMPs are non-relativistic, and that their local density is 0.3\,GeV/cm$^3$ \cite{Green:2017odb}, the interaction rate in a direct detection experiment can be expressed as~\cite{Pospelov:2008jk}:

\begin{equation}
\label{eq:vrate}
R\simeq\frac{4\times10^{23}}{A}\frac{\alpha'}{\alpha}\left(\frac{\rm keV}{m_{v}}\right)\left(\frac{\sigma_{pe}}{\rm b}\right)\rm{kg^{-1}d^{-1}}, 
\end{equation}

\noindent
and 

\begin{equation}
\label{eq:arate}
R\simeq\frac{1.29\times10^{19}}{A}g_{ae}^{2}\left(\frac{m_{a}}{\rm keV}\right)\left(\frac{\sigma_{pe}}{\rm b}\right) \rm{kg^{-1}d^{-1}},
  \end{equation}
  
\noindent
for  vector  and pseudo-scalar super-WIMPs, respectively. $g_{ae}=2m_{e}f_{a}^{-1}$ is the axio-electric coupling, $m_e$ is the mass of the electron, and $A$ is the atomic number of the target atom. 

\begin{figure}[h!]
\centerline{\includegraphics[width=.5\textwidth]{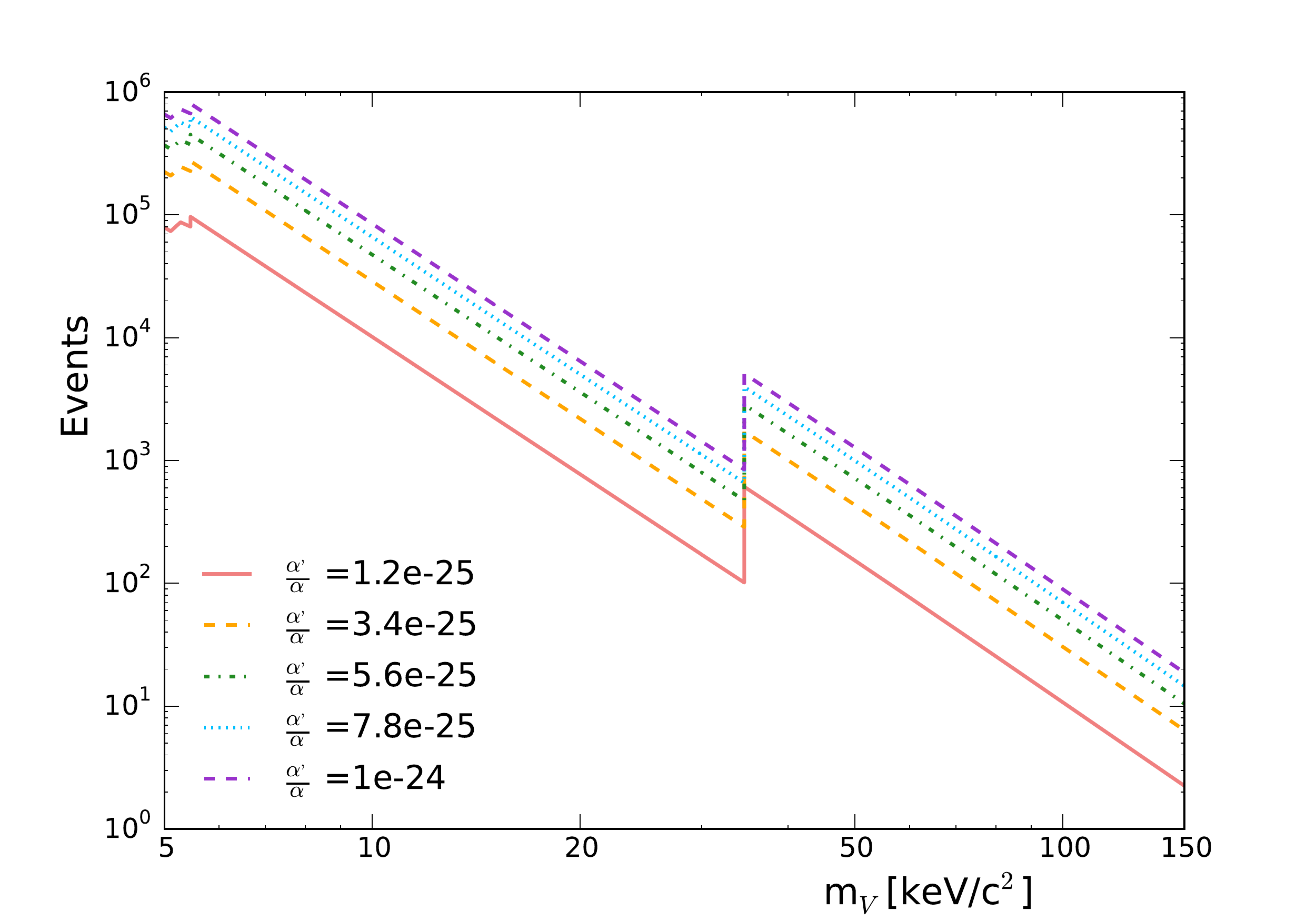}}
\caption{Predicted total number of events, assuming an infinite energy resolution, for vector super-WIMPs in XENON100 as a function of mass for a range of coupling constants. The assumed exposure is 34\,kg$\times$225 live days of data.}
\label{fig:vcount}
\end{figure}

\begin{figure}[h!]
\centerline{\includegraphics[width=.5\textwidth]{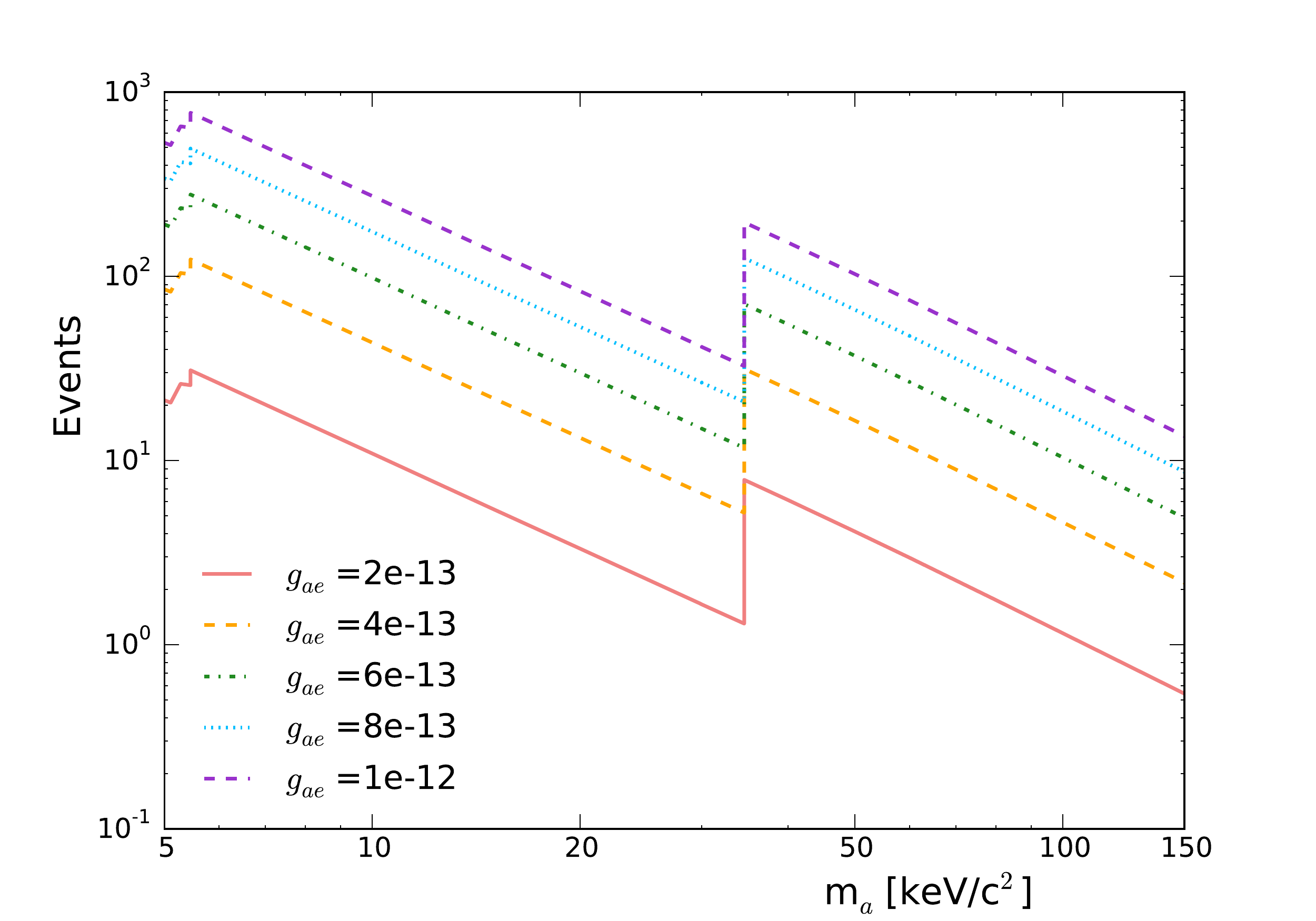}}
\caption{Predicted total number of events, assuming an infinite energy resolution, for pseudo-scalar super-WIMPs in XENON100 as a function of mass for a range of coupling constants. The assumed exposure is 34\,kg$\times$225 live days of data.}
\label{fig:acount}
\end{figure}

With equations \ref{eq:vrate} and \ref{eq:arate}, we can predict the interaction rate in XENON100, assuming a range of coupling constants, and an exposure of 34\,kg$\times$224.6\,live days, as shown in figures \ref{fig:vcount} and \ref{fig:acount} for  vector and pseudo-scalar super-WIMPs, respectively. The structures observed around 35\,keV, as well as at lower energies are due to an increase in the cross section of the photoelectric effect in xenon at these energies, when new atomic energy levels are excited~\cite{Veigele:1973aw}.

\section{\label{sec:analysis} Data Analysis}

We perform this analysis using XENON100 Run\,II science data, with 224.6\,live days of data and 34\,kg of LXe in the fiducial region. Our data selection and treatment is similar to the one described in~\cite{Aprile:2014eoa}, with a few differences, to be detailed below.  In particular, this is the first analysis of the XENON100 electronic recoil data that extends up to an energy of 140\,keV. The overall background in the low-energy region is 5.3$\times$10$^{-3}$\,events/(kg\,d\,keV) and it is flat in shape, because it is dominated by Compton scatters of high-energy gammas originating from the radio-activity of detector materials~\cite{Aprile:2011vb}. 

\subsection{Event selection and signal region}

An interaction in the LXe-TPC gives rise to an S1 and a correlated S2 signal with a certain number of photoelectrons (PE) observed by the photosensors.  To select valid events, we apply event selection criteria as described in~\cite{Aprile:2014eoa}. These include basic data quality selection criteria, single-scatter and fiducial volume selection, and signal consistency checks. Furthermore, we use position-corrected S1 and S2 quantities as detailed in~\cite{Aprile:2011dd} and \cite{Aprile:2012vw}.  In contrast with~\cite{Aprile:2014eoa}, we determine the energy deposition of an interaction in LXe by  using  a linear combination of the prompt and delayed scintillation light signals, where we employ the S2 signal detected by the bottom PMT arrays only (S2$_b$), due to smaller required corrections. This energy scale improves our energy resolution (at 1-$\sigma$) from  14\%, based on S1-only, to 5.7\% at an energy of $\sim$100\,keV~\cite{Aprile:2011dd}.

\begin{figure}[h!]
\centerline{\includegraphics[width=.5\textwidth]{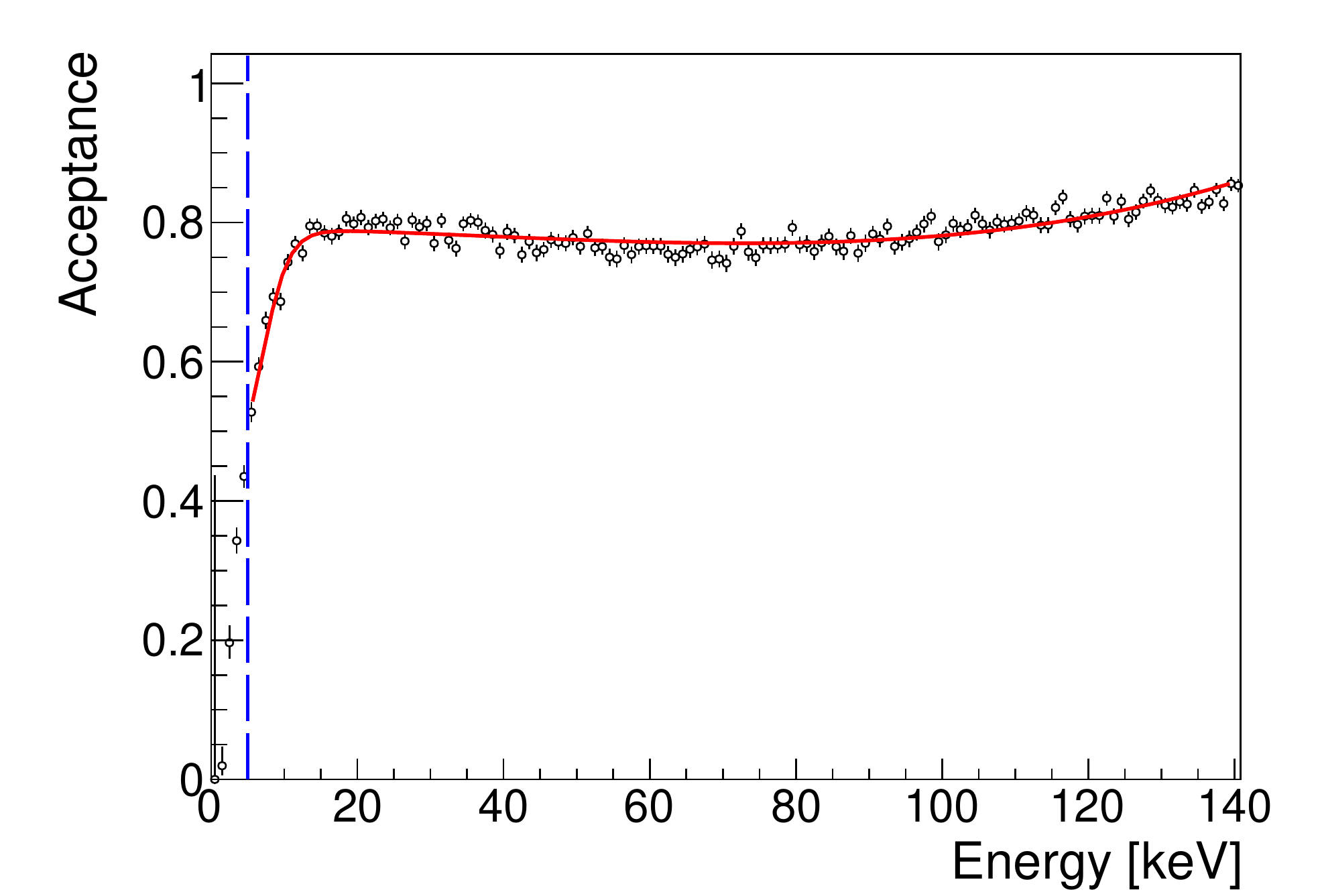}}
\caption{Calculated acceptance $\epsilon$ (black dots) of all the event selection criteria as a function of energy, along with a fit (red curve) to the distribution.  The energy threshold employed in this analysis (5\,keV) is shown by the vertical dashed line.  }
\label{fig:totacc}
\end{figure}

The combined acceptance of all applied event selection criteria, evaluated on calibration data acquired with  $^{60}$Co and $^{232}$Th sources, is shown in figure\,\ref{fig:totacc}. This acceptance, around 80\%, is rather flat in energy above $\sim$15\,keV, below which it decreases to reach 53\% at 5\,keV.

\begin{figure}[!!!h]
\centerline{\includegraphics[width=.5\textwidth]{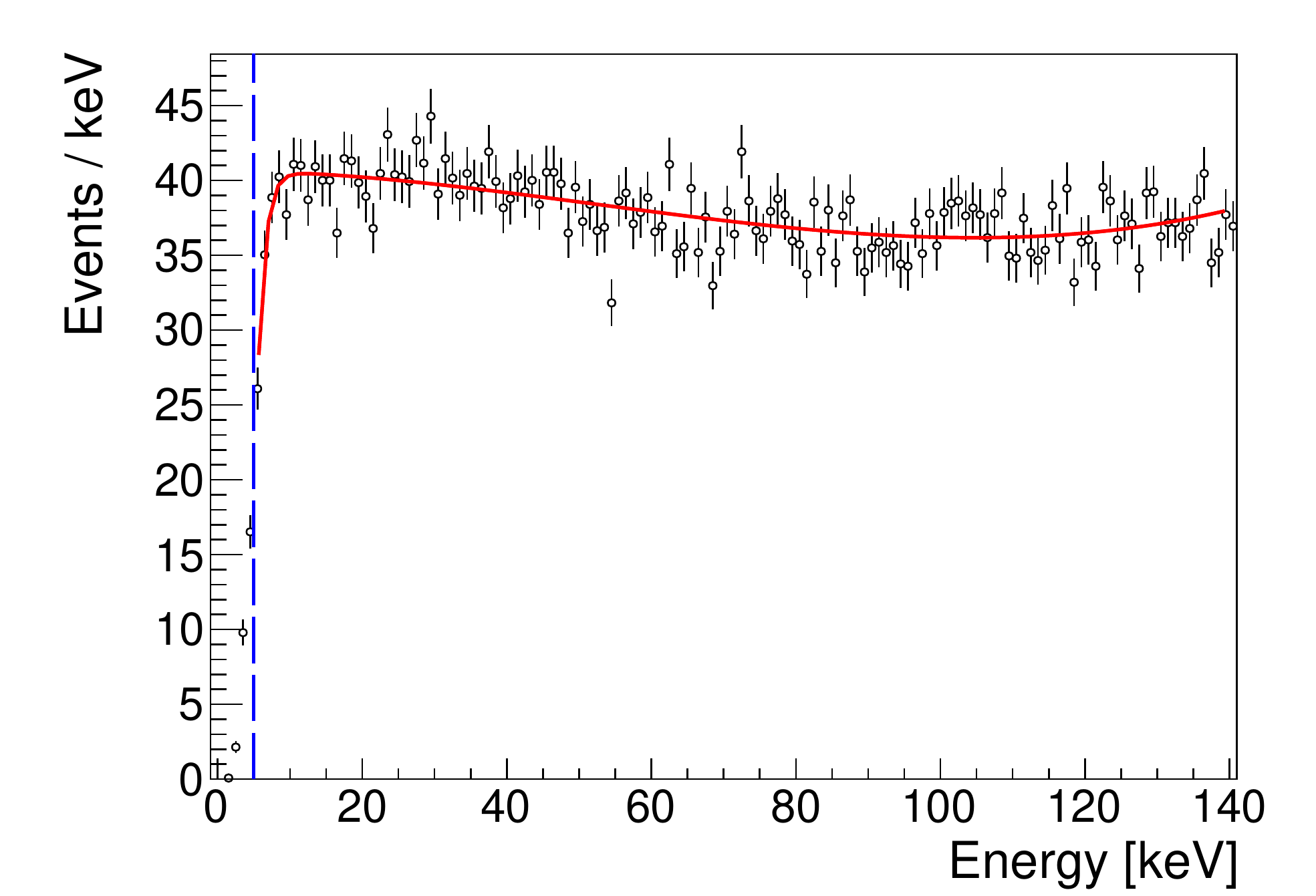}}
\caption{Background model $N_b \times f_b$ (red line) scaled to the correct exposure. The model is based on $^{60}$Co and $^{232}$Th calibration data (black dots) and used in Eq.\,\ref{eq:likelihood}. The energy threshold of 5\,keV  is shown by the vertical dashed line.}
\label{fig:bgmodel}
\end{figure}

The background events in the region of interest are predominantly due to interactions of $\gamma$-rays from decays of radioactive isotopes in the detector materials, yielding low-energy Compton-scatters, and from $\beta$-decays of the $^{222}$Rn and $^{85}$Kr isotopes distributed in the liquid xenon~\cite{Aprile:2011vb}. To model the expected shape of the background distribution in the region of interest, we employ calibration data acquired with  $^{60}$Co and $^{232}$Th sources. The spectral shape is parameterised with a modified Fermi function, as shown in figure\,\ref{fig:bgmodel}, together with the calibration data.

\subsection{\label{sec:stats} Statistical Method}

A profile likelihood analysis is employed to constrain the coupling constants $g_{ae}$ and $\alpha'/\alpha$, as described in \cite{Cowan:2010js,Aprile:2011hx}. The full likelihood function is given by

\begin{equation}
\label{eq:likelihood}
\begin{split}
\mathcal{L}((g_{ae}, \alpha'/\alpha), N_{b}) = {\rm{Poiss}}(N| N_s + N_b)  \\
\prod_{i=1}^{N} \frac{N_s f_s(E_{i}) + N_b f_b(E_{i})}{N_s+N_b},  
 \end{split}
\end{equation}

\noindent
where the parameters of interest are  $g_{ae}$ or $\alpha'/\alpha$ and $N_{b}$ and $N_{s}$ are the expected number of background and signal events in the search region, respectively. $N_{b}$ is considered as nuisance parameter,  $N_{s}$  is a function of the coupling constant, and the functions $f_b$ and $f_s$ are the  background and signal probability distribution functions. $N$ is the total number of observed events and $E_{i}$ corresponds to the energy of the $i$th event. We model the expected event rate at a given energy with a Gaussian smeared with the energy resolution of the detector and multiplied by the total acceptance of the data selection criteria, shown in figure\,\ref{fig:totacc}. 

Our selected region of interest extends from 5\,keV to 140\,keV. The lower bound is chosen such as to have $>$50\% acceptance of the data selection criteria, as mentioned above.  The upper bound is given by the width of the 164\,keV line from $^{131\mathrm{m}}$Xe decays, present in both calibration and physics data due to the activation of xenon during an AmBe neutron calibration. The relative resolution at this energy is 4.6\%, and we consider the region up to 3-$\sigma$ to the left of the peak.

For each considered mass  $m_a$ of the bosonic super-WIMP, the $\pm$2-$\sigma$-region, as determined by the energy resolution at the given energy, is blinded. The number of events outside of this region is then used to scale the background model $f_b$ shown in figure\,\ref{fig:bgmodel}, and thus to predict the number of expected background events in the signal region. Since the $\pm$2-$\sigma$ window must not exceed the search region between $(5-140)$\,keV, the range of bosonic super-WIMP masses is restricted to $(8-125)$\,keV. We use the CL$_s$ prescription~\cite{Read:2000ru} to protect against overly constrained parameters due to downward fluctuations in the background.

\section{\label{sec:results} Results}

\begin{figure}[h!]
\centerline{\includegraphics[width=.52\textwidth]{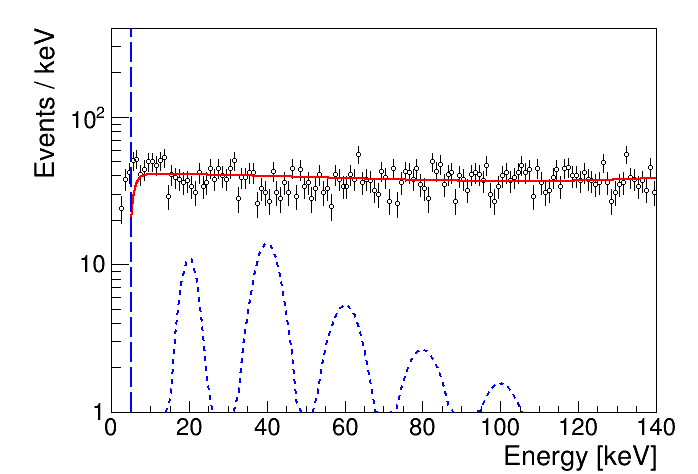}}
\caption{Distribution of events (black dots with error bars) in the super-WIMP search region between $(5-140)$\,keV. The background model (red curve)  along with the expected signal for various pseudo-scalar super-WIMP masses (20, 40, 60, 80 and 100\,keV) and a coupling of $g_{Ae} = 1\times10^{-12}$  (blue dashed peaks) is also shown.  The vertical dashed line indicated the energy threshold in this analysis.}
\label{fig:signalexp}
\end{figure}

Figure\,\ref{fig:signalexp} shows the distribution of events in the region of interest, together with the expected signal for different pseudo-scalar super-WIMP masses and an assumed coupling of $g_{ae} = 1\times10^{-12}$. We also assume that these particles constitute all the dark matter in our galaxy, with a local density of 0.3\,GeV/cm$^3$. The widths of the mono-energetic signals are given by the energy resolution of the detector, in combined energy scale, at the given S1 and S2$_b$ signal size. 

Our data are compatible with the background only hypothesis, and no excess above the predicted background is observed.

\begin{figure}[h!]
\centering
\includegraphics[width=.5\textwidth]{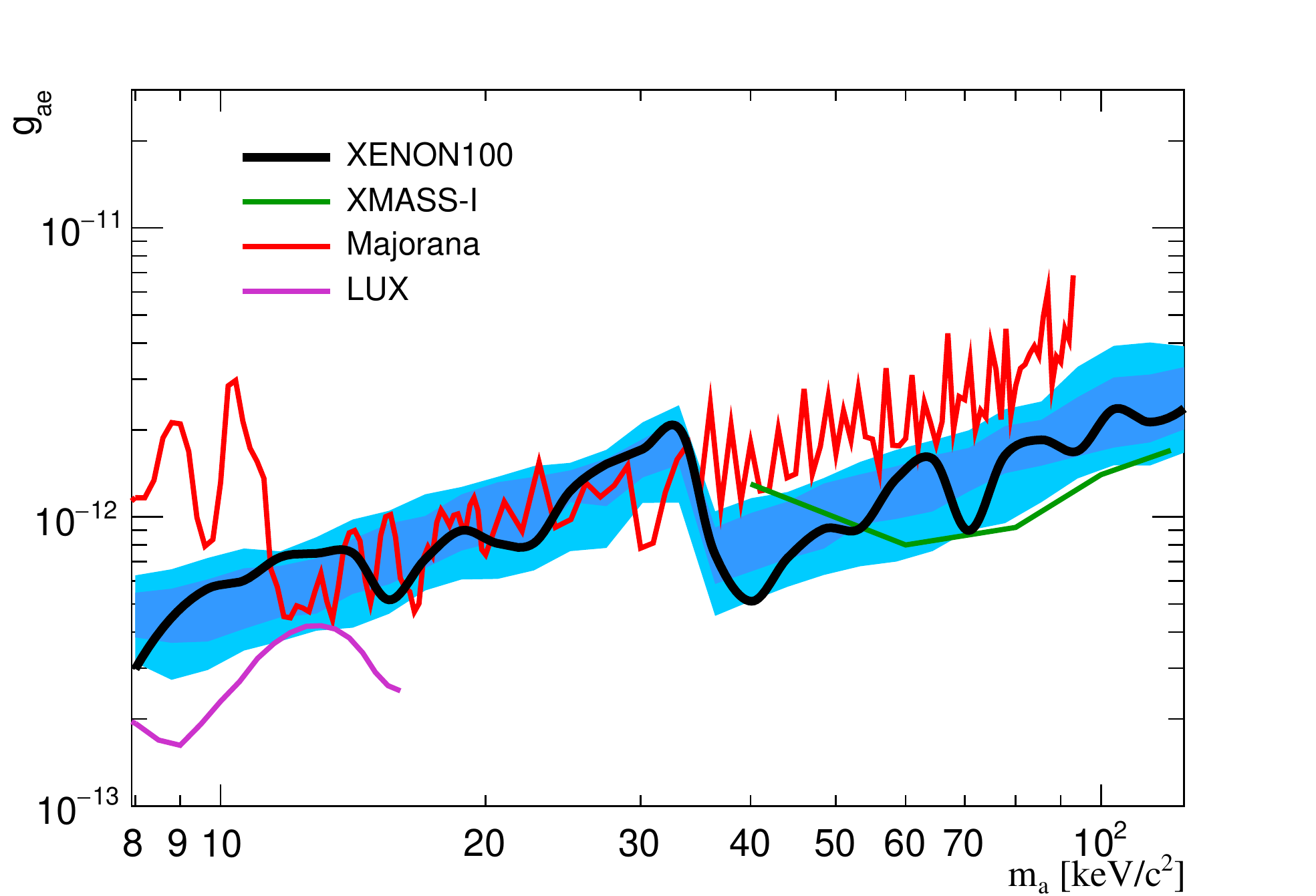}
\includegraphics[width=.5\textwidth]{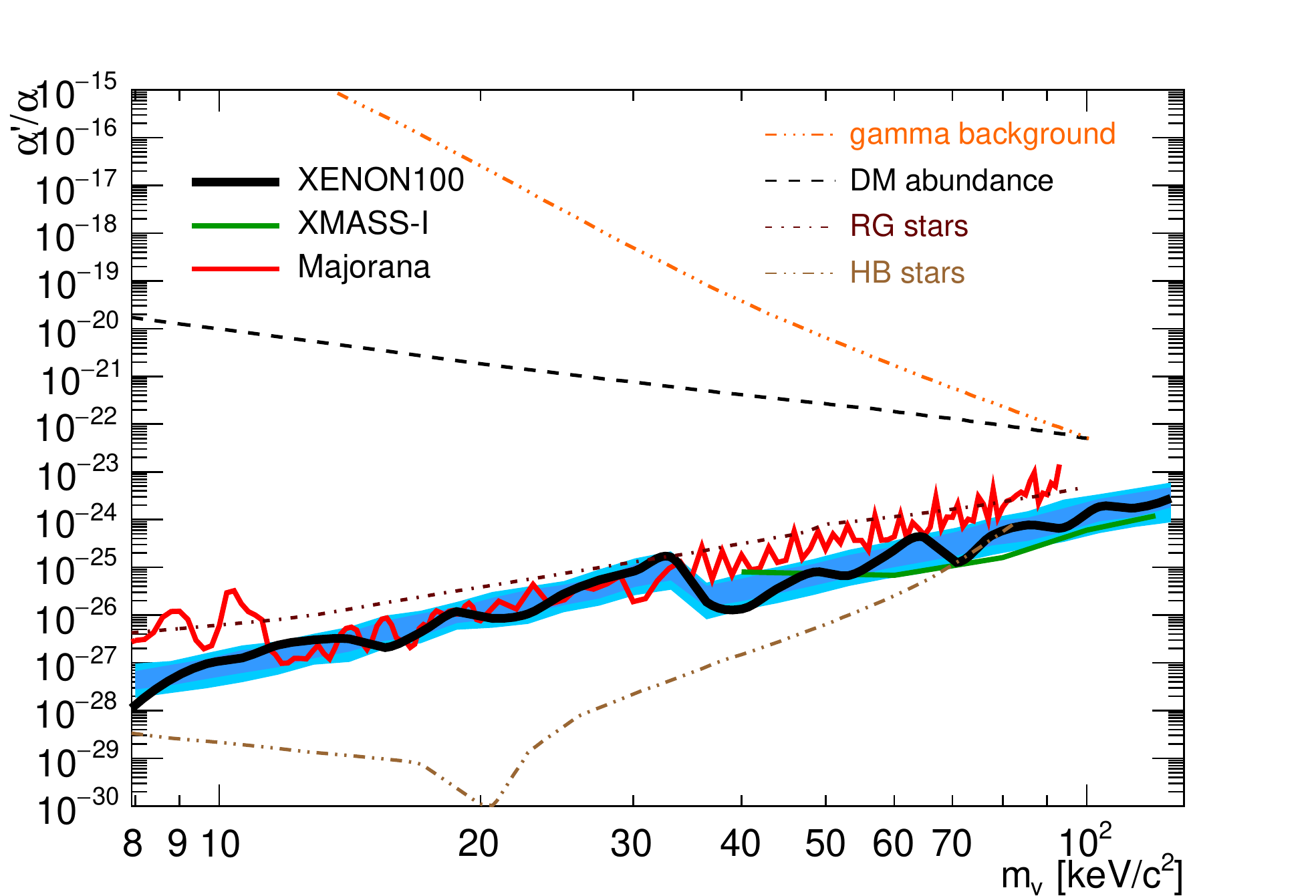}
\caption{The XENON100 upper limits, at 90\% C.L., on the coupling of pseudo-scalar (top) and vector (bottom) super-WIMPs as a function of particle mass. The 1-$\sigma$ (2-$\sigma$) expected sensitivity is shown by the dark (light) blue bands. We compare our limits to the results obtained by the XMASS-I~\cite{Abe:2014zcd} (green), LUX~\cite{Akerib:2017uem}  (magenta) and Majorana Demonstrator~\cite{Abgrall:2016tnn} (red) experiments, as well as with astrophysical constraints from  the gamma background, the dark matter abundance, red giant stars  and horizontal branch stars (dashed and dash-dotted) \cite{Pospelov:2008jk} for the case of the vector super-WIMP.}
\label{fig:pllimit}
\end{figure}

Figure\,\ref{fig:pllimit} shows the 90\% C.L. exclusion limit for pseudo-scalar (top panel) and vector (bottom panel) super-WIMPs. Our 1-$\sigma$ (2-$\sigma$) expected sensitivity is shown by the dark (light) blue bands. The step in sensitivity around 35\,keV/c$^2$ is due to an increase in the photoelectric cross section as new atomic energy levels are excited. We compare these results with those obtained by the XMASS-I~\cite{Abe:2014zcd}, LUX~\cite{Akerib:2017uem} and the Majorana Demonstrator~\cite{Abgrall:2016tnn} experiments. While the XMASS-I constraints are more stringent in the mass region above 50\,keV/c$^2$, our results improve upon these at lower masses, and in particular extend the excluded mass region down to 8\,keV/c$^2$. Our results are comparable to those of the Majorana Demonstrator in the mass range $\sim(12-32)$\,keV/c$^2$, but more stringent at higher masses. Finally, LUX presents the most constraining limit at masses of the pseudoscalar particle below 16\,keV/c$^2$.

The observed fluctuations in our limit are caused by statistical fluctuations in the background (see figure~\ref{fig:signalexp}). Due to the expected mono-energetic shape of the signal, the limit is very sensitive to such fluctuations. We have also studied the impact of systematic uncertainties on the analysis. In particular, we have considered the impact of varying the overall event selection acceptance, the energy resolution, as well as the energy scale. The combined effect of all systematic uncertainties changes the final result by around 10\%, however this contribution is small compared to our statistical uncertainty that is accounted for in the profile likelihood analysis.

XENON100 thus sets new and stringent upper limits in the $(8-125)$\,keV/c$^2$ mass range. At 90\% C.L. it excludes couplings to electrons $g_{ae} > 3\times10^{-13}$  for pseudo-scalar  super-WIMPs and $\alpha'/\alpha > 2\times10^{-28}$ for vector  super-WIMPs. These limits are derived under the assumption that super-WIMPs constitute all of the galactic dark matter. 

\vspace*{1cm}

\section*{Acknowledgements}
We gratefully acknowledge support from the National Science Foundation, Swiss National Science Foundation, Deutsche Forschungsgemeinschaft, 
Max Planck Gesellschaft, German Ministry for Education and Research, Netherlands Organisation for Scientific Research, Weizmann Institute of Science, 
I-CORE, Initial Training Network Invisibles (Marie Curie Actions, PITNGA-2011-289442), Fundacao para a Ciencia e a Tecnologia, Region des Pays de la Loire, 
Knut and Alice Wallenberg Foundation, Kavli Foundation, and Istituto Nazionale di Fisica Nucleare. 
We are grateful to Laboratori Nazionali del Gran Sasso for hosting and supporting the XENON project.

\bibliography{superwimp_paper_v07}

\end{document}